\documentclass[12pt]{article}
\usepackage{epsfig}

\begin{document}

\begin{titlepage}

\hfill FTUV  08-0815

%\hfill IFIC

\vspace{1.5cm}

\begin{center}
\ \\
{\bf\large Monopolium production from photon fusion at the Large
Hadron Collider}
\\
\date{ }
\vskip 0.70cm

Luis N. Epele$^{a}$, Huner Fanchiotti$^{a}$, Carlos A. Garc\'{\i}a
Canal$^{a}$
\\ and Vicente Vento$^{b,c}$

\vskip 0.30cm

{(a) \it Laboratorio de F\'{\i}sica Te\'{o}rica, Departamento de
F\'{\i}sica, IFLP \\ Facultad de Ciencias Exactas, Universidad
Nacional de La Plata
\\C.C. 67, 1900 La Plata, Argentina.}\\({\small
E-mail: epele@fisica.unlp.edu.ar, garcia@fisica.unlp.edu.ar,
huner@fisica.unlp.edu.ar})

\vskip 0.3cm

{(b) \it Departamento de F\'{\i}sica Te\'orica and Instituto de
F\'{\i}sica
Corpuscular}\\
{\it Universidad de Valencia and Consejo Superior
de Investigaciones Cient\'{\i}ficas}\\
{\it E-46100 Burjassot (Valencia), Spain} \\ ({\small E-mail:
Vicente.Vento@uv.es}) \vskip 0.3cm {(c) \it TH-Division, PH
Department, CERN} \\
{\it CH-1211 Gen\`eve 23, Switzerland}
\end{center}

\vskip 1cm \centerline{\bf Abstract}

Magnetic monopoles have attracted the attention of physicists since
the founding of the electromagnetic theory. Their search has been a
constant endeavor which was intensified when Dirac established the
relation between the existence of monopoles and charge quantization.
However, these searches have been unsuccessful. We have recently
proposed that monopolium, a monopole-antimonopole bound state, so
strongly bound that it has a relatively small mass, could be easier
to find and become an indirect but clear signature for the existence
of magnetic monopoles. In here we extend our previous analysis for
its production to two photon fusion at LHC energies.

 \vspace{1cm}

\noindent Pacs: 14.80.Hv, 95.30.Cq, 98.70.-f, 98.80.-k


\end{titlepage}

\section{Introduction}

The theoretical justification for the existence of classical
magnetic poles is that they add symmetry to Maxwell's equations and
explain charge quantization
\cite{Dirac:1931kp,Dirac:1948um,Jackson:1982ce} . Dirac formulated
his theory of monopoles considering them point-like particles and
quantum mechanical consistency conditions lead to the so called
Dirac Quantization Condition (DQC),

\begin{equation} e \, g = \frac{N}{2} \;, \mbox{  N = 1,2,...}\;
, \end{equation}

\noindent where $e$ is the electron charge, $g$ the monopole
magnetic charge and we use natural units $\hbar = c =1$.

Numerous experimental searches for monopoles have been carried out
but all have met with
failure\cite{Craigie:1986ws,Giacomelli:2005xz,Martin:1989ms,
Milton:2006cp,Eidelman:2004wy,Yao:2006px,Mulhearn:2004kw}. The last,
carried out by the CDF collaboration at the Fermi National
Laboratory \cite{Abulencia:2005hb}, found no monopoles and
established a lower mass limit of 360 GeV.

This lack of experimental confirmation has led many physicists to
abandon the hope in their existence. A way out of this impasse is
the old idea of Dirac \cite{Dirac:1931kp,Zeldovich:1978wj}, namely,
monopoles are not seen freely because they are confined by their
strong magnetic forces forming a bound state called monopolium
\cite{Hill:1982iq,Dubrovich:2002gp}. This idea was the leitmotiv
behind our recent research \cite{Epele:2007ic}, namely we proposed
that monopolium  might be easier to detect than free monopoles. We
showed that certain parameterizations of the mass and the width,
allowed for such a scenario.

The Large Hadron Collider (LHC) will soon enter in operation and
will probe the new energy frontier opening possibilities for new
physics including the discovery of magnetic monopoles either
directly, a possibility contemplated long time ago,
\cite{Ginzburg:1981vm,Ginzburg:1982fk}, or through the discovery of
monopolium, as we have been advocating. This development motivates
our present research which analyzes the production of monopolium at
LHC by the mechanism of photon fusion.

Recently, Dougall and Wick have presented a calculation of
monopole--antimonopole production from photon fusion at proton
colliders \cite{Dougall:2007tt,Dougall:2007zz}. The mass limit of
CDF was obtained assuming Drell-Yan (DY) production which dominates
over other processes at Tevatron energies. Dougall and Wick have
shown that at LHC energies photon fusion is the dominant process
\cite{Dougall:2007tt,Dougall:2007zz}. We proceed to calculate the
production of monopolium by means of this process and compare our
results with monopole--antimonopole production. We conclude that, if
monopolium is a strongly bound state, its cross section is larger
than that for creating a heavy monopole-antimonopole pair and
consequently gives rise to a more clear experimental signal.

\section{Monopolium production}

A useful computational theory of monopoles does not currently exist
to perform a direct production calculation. For this reason we will
employ a minimal model of monopole interaction which assumes a
monopole photon-coupling which is proportional to the monopole's
induced electric field $g\beta$ for a monopole moving with velocity
$\beta$ \cite{Dougall:2007tt,Dougall:2007zz,Kalbfleisch:2000iz}.
This approximation can be shown to be almost equivalent to the low
energy effective theory of Ginzburg and Schiller
\cite{Ginzburg:1998vb,Ginzburg:1999ej}. This theory was derived from
the standard electroweak theory in the one loop approximation
leading to an effective coupling  proportional to $ g_{eff} \sim
\frac{\omega}{m}\, g$, where $\omega$ is a kinematical energy scale
of the process which is below the monopole production threshold,
thus rendering the theory perturbative. This is so because that in a
photon fusion diagram the dynamical scale is $\sqrt{E^2-4m^2}$, thus

\begin{equation}
\frac{\omega}{m} \sim \frac{\sqrt{E^2-4m^2}}{2m}\sim \frac{E \beta
}{2m},
\end{equation}
and therefore if $E \sim 2m$, i.e., kinetic terms are small, both
schemes coincide. Here $\omega$ describes the active energy scale, E
the center of mass energy, $m$ the monopole mass and $\beta$ the
velocity.

The Dirac quantization condition does not specify the spin of the
monopoles. We choose here monopoles of spin 1/2, following Dougall
and Wick \cite{Dougall:2007tt,Dougall:2007zz}, coupled in monopolium
to spin $0$ in order to have a minimum energy radial structure.

\begin{figure}
\centerline{\epsfig{file=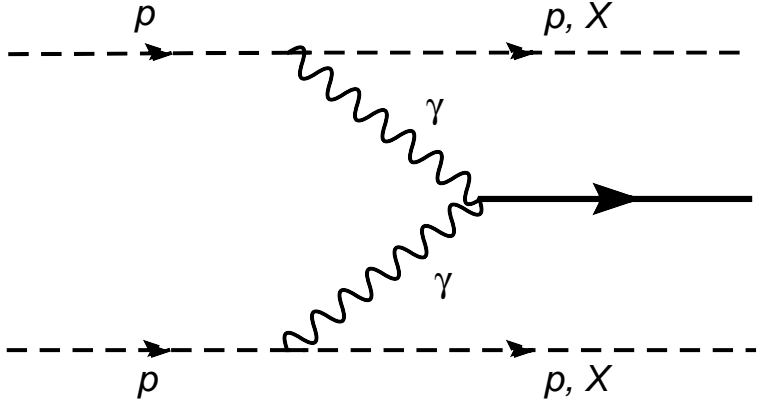,width=10cm,angle=0}}
\caption{\small{Diagrammatic description of the reaction studied.}}
\label{pXpXM}\end{figure}

We then study the expected production process at LHC, namely

\begin{eqnarray}
p + p &  \rightarrow &  p(X) + p(X) + M,
\end{eqnarray}
shown in Fig.\ref{pXpXM}, where $p$ represents the proton, $X$ an
unknown final state and $M$ the monopolium. This diagram summarizes
the three possible processes:

\begin{itemize}
\item [i)] inelastic $p+ p \rightarrow X+X +(\gamma \gamma)
\rightarrow X + X + M$
\item [ii)] semi-elastic $ p + p \rightarrow p + X + (\gamma \gamma)
\rightarrow p + X + M$
\item [iii)] elastic $p + p \rightarrow p + p + (\gamma \gamma)
\rightarrow p + p + M$.
\end{itemize}

In the inelastic scattering, both intermediate photons are radiated
from partons (quarks or  antiquarks) in the colliding protons.

In the semi-elastic scattering one intermediate photon is radiated
by a quark (or antiquark), as in the inelastic process, while the
second photon is radiated from the other proton, coupling to the
total proton charge and leaving a final state proton intact.

In the elastic scattering both intermediate photons are radiated
from the interacting protons leaving both protons intact in the
final state.

The full $\gamma \gamma$ calculation includes contributions from
these three individual regimes.

\begin{figure}
\centerline{\epsfig{file=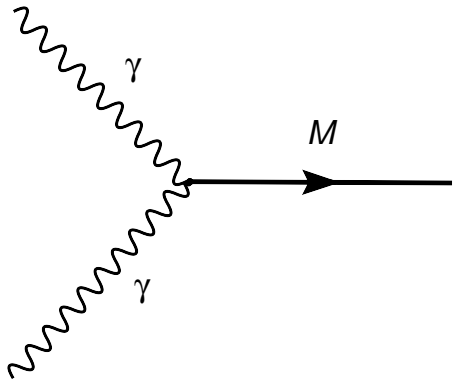,width=6cm,angle=0}}
\caption{\small{Diagrammatic description of the elementary
subprocess of the monopolium production from photon fusion.}}
\label{ggM}\end{figure}

We next proceed to describe the elementary subprocess shown in Fig.
\ref{ggM}, which deals only with photons and monopolium, and will
come back to the full $p p$ scattering treatment later on. The
standard expression for the cross section of this elementary
subprocess results in

\begin{equation}
\sigma (2 \gamma \rightarrow M) = \frac{4\pi}{E^2}  \frac{M ^2
\,\Gamma (E) \, \Gamma_M}{\left(E^2 - M^2\right)^2 +
M^2\,\Gamma_M^2} \label{ppM}
\end{equation}
where we have assumed that monopolium decays with a width $\Gamma_M$
and $\Gamma (E)$, with $E$ off mass shell, describes the production
cross section. The width $\Gamma_M$ arises from the softening of a
delta function, $\delta(E^2 - M^2)$ and therefore is, in principle,
independent of the production rate $\Gamma (E)$
\cite{Peskin:1995hc}.

We enter now the computation of the $\Gamma (E)$, which represents
the width of the $2 \gamma$ decay of monopolium. The calculation,
following standard field theoretic techniques of the decay of a
non-relativistic bound state \cite{Peskin:1995hc,Jauch:1975sp},
leads to

\begin{equation}
\Gamma (E) =
\frac{32\,\pi\,\alpha_g^2}{M^2}\,\left|\psi_M(0)\right|^2 .
\end{equation}
We have used the conventional approximations, namely that the
monopoles  are almost on shell and that in the calculation of the
elementary process we have neglected the binding energy, i.e. $M =
2m$. However, we express the final formula in terms of the
monopolium mass, $M$, because the latter will take care of the
binding. Here $\alpha_g$ corresponds to the photon--monopole
coupling and $\psi_M$ is the monopolium ground state wave function.

Using the Coulomb wave functions of ref.\cite{Epele:2007ic}
expressed in the most convenient way to avoid details of the
interaction, which will be parameterized by the binding energy, one
has

\begin{equation}
|\psi_{M}(0)|^2 = \frac{1}{\pi}(2 - \frac{M}{m})^{3/2}\; m^3,
\end{equation}
and the scheme of Dougall and Wick
\cite{Dougall:2007tt,Dougall:2007zz} adapted to monopolium
production, gives rise to

\begin{equation}
\Gamma(E)= \frac{2 \beta^4}{M^2 \alpha^2} (2- \frac{M}{m})^{3/2}
m^3.
\end{equation}
Here, $\alpha$ is the fine structure constant and $\beta$ the
monopolium velocity,

\begin{equation}
\beta= \sqrt{1-\frac{M^2}{E^2}}.
\end{equation}
which is the velocity of the monopoles moving in the monopolium
system.

Note that due to the value of $\beta$ the width vanishes at the
monopolium mass, where the velocity is zero. Therefore a static
monopolium is stable under this interaction.

\begin{figure}[t]
\centerline{\epsfig{file=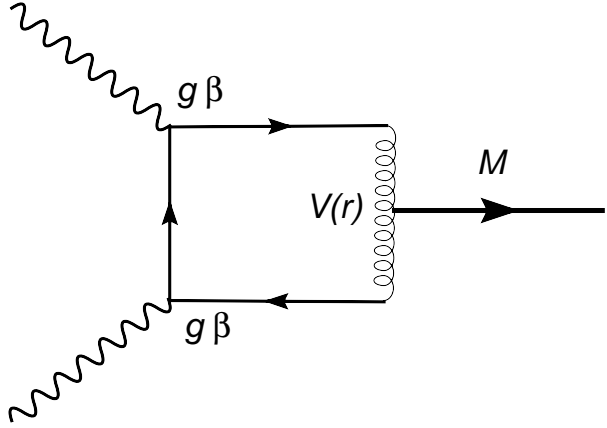,width=8cm,angle=0}}
\caption{\small{Diagramatic representation of the model describing
the coupling of photons to monopolium.}} \label{gbetab}\end{figure}

A caveat must be made. There is a duality of treatments in the above
formulation, see Fig. \ref{gbetab}. The static coupling is treated
as a Coloumb like interaction of coupling $g$ binding the monopoles
into monopolium, although ultimately the details are eliminated in
favor of the binding energy parameterized by the monopolium mass
$M$. We find in this way a simple parametric description of the
bound state. The dynamics of the production of the virtual
monopoles, to be bound in monopolium, is described in accordance
with the effective theory \cite{Dougall:2007tt,Dougall:2007zz}, and
this coupling is $\beta g$. This is similar to what is done in heavy
quark physics \cite{Pennington:2005ww}(see his figure 5), where the
wave function is obtained by a parametric description using
approximate strong dynamics while the coupling to photons is
elementary.

The production cross section can now be written as

\begin{equation}
M^2 \sigma(2\gamma \rightarrow M) = \frac{2\sqrt{2} \pi R^{3/2}
(R-1)^{1.5}}{\alpha^2 {\cal E}^6}\frac{\bar{\Gamma}_M ({\cal E}^2
-1)^2}{({\cal E}^2 -1)^2 +  \bar{ \Gamma}_M^2},
\label{ggxsecM}\end{equation}
where $R=2m/M , (1<R< \infty)$. This parameter ratio describes the
binding energy of monopolium in units of $M$, since $M= 2m +
E_{binding}$. ${\cal E} = E/M$ is the center of mass energy measured
in units of $M$ and $\bar{ \Gamma}_M = \Gamma_M/M$ is the decay
width of monopolium also measured in units of $M$. The right hand
side is adimensional and therefore the above expression gives the
cross section in units of $1/M^2$. The monopolium width, which
vanishes with our dynamics, arises from higher order effects or
other possible dynamics and we consider it as a parameter. Even in
the case monopolium would be stable at rest the width would
parameterize the beam width \cite{Peskin:1995hc}.

\begin{figure}
\centerline{\epsfig{file=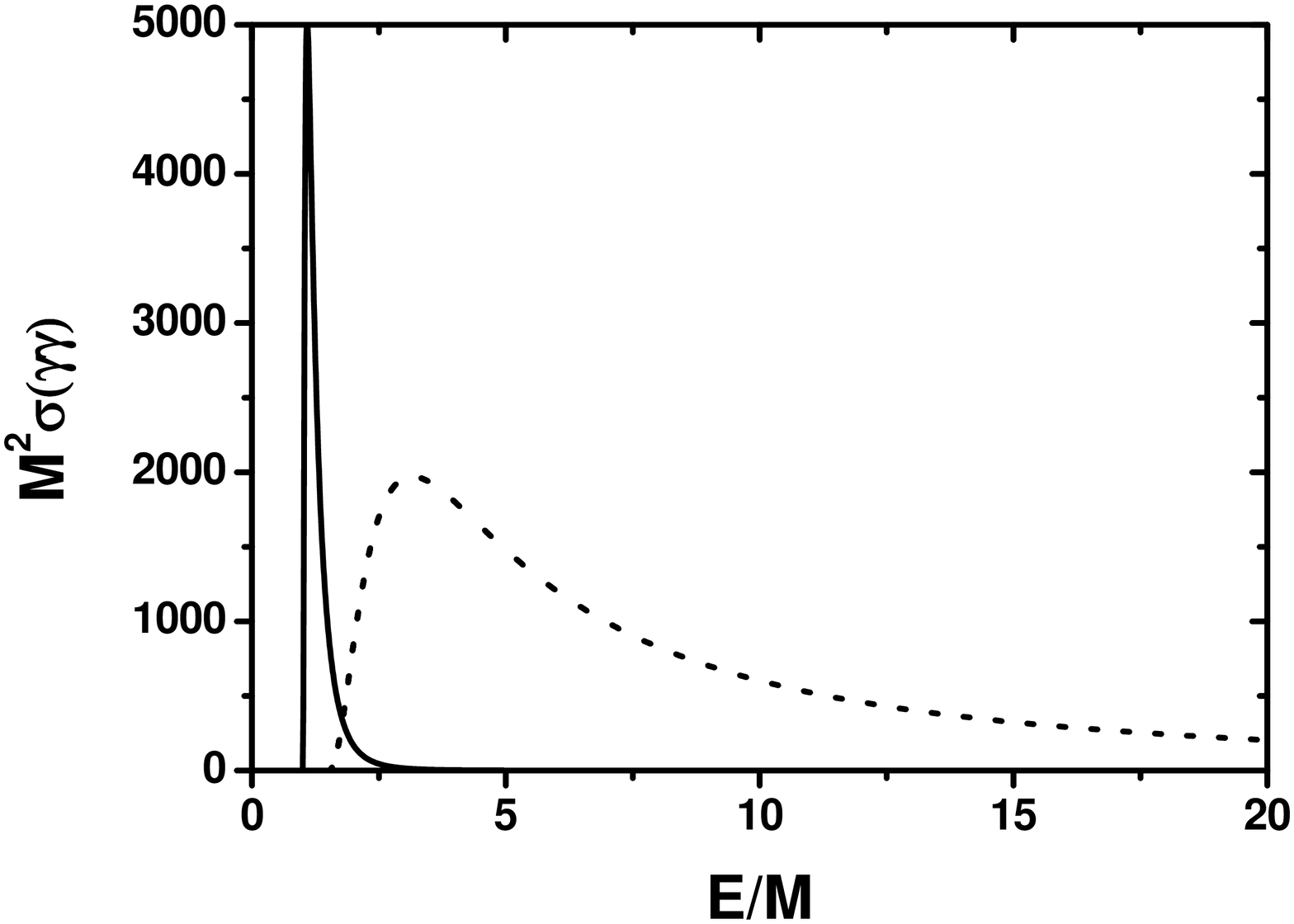,width=8cm,angle=270}}
\caption{\small{We show the photon fusion production cross section
for monopolium (solid curve) for $R=1.5$ and $\bar{\Gamma}_M =0.1$,
and for monopole-antimonopole (dotted curve)  as a function of the
energy variable ${\cal E} = E/M$.}} \label{ggxsec}\end{figure}

In Fig. \ref{ggxsec} we show the photon fusion cross section for
monopolium for a value of $R = 1.5$ and $\bar{\Gamma}_M = 0.1$,
together with that for monopole-antimonopole obtained in ref.
\cite{Dougall:2007tt,Dougall:2007zz}, which we have simply rewritten
using $M$ as the energy unit,

\begin{equation}
M^2 \sigma(2\gamma \rightarrow M)= \frac{\pi\,
R^2\,(1-\beta'^2)\beta'^5}{14 \alpha^2 m^2}
\left(\frac{3-\beta'^4}{2 \beta'}
\log\left({\frac{1+\beta'}{1-\beta'}}\right) -(2-\beta^2)\right),
\label{ggxsecmm}
\end{equation}
where $\beta'=\sqrt{1-\frac{4m^2}{E^2}}$.

.

The qualitative features of both cross sections are radically
different. The monopolium cross section is a spike slightly above
the monopolium mass, where it vanishes, assuming a reasonable width
$\Gamma_M < M$, while the monopole-antimonopole cross section is a
soft curve extending over a large energy region. It is also clear
from Eq. (\ref{ggxsecM}) that the height of the pick strongly
depends on the value of the binding, i.e., the larger the binding
energy the larger will be $R$ and the higher will be the pick.

It is important to note that the approach followed to describe the
coupling ($g\rightarrow g\beta$) provides us with a large negative
power of energy which makes the monopolium cross section fall off
very rapidly. In the Ginzburg-Schiller approach, four powers of the
energy are substituted by four powers of the monopolium mass and
therefore one obtains a larger effective width for the pick, and
thus a larger integrated cross section for the same values of R and
$\bar{\Gamma}_M$.

\section{Cross section estimates}

We calculate $\gamma \gamma$ fusion for monopolium production
following the formalism of Drees et al. \cite{Drees:1994zx}
benefitting from the the full documentation of the calculation in
the work of Dougall and Wick \cite{Dougall:2007tt,Dougall:2007zz}.
We obtain therewith the $p p$ cross section for monopolium and for
monopole-antimonopole production within the same computational
codes.

The full $p p$ calculation includes contributions of three types:
inelastic, semi--elastic, and elastic scattering. We sum these
individual contributions to find the total $pp$ cross--section,
$\sigma_{tot}$.

In the inelastic scattering, $p + p\rightarrow X+ X + (\gamma
\gamma) \rightarrow X +X + M$, to approximate the quark distribution
within the proton we use the Cteq6--1L parton distribution functions
\cite{CTEQ} and choose $Q^2=\hat{s}/4$ throughout.

We employ an equivalent--photon approximation for the photon
spectrum of the intermediate quarks \cite{Williams:1934ad,von
Weizsacker:1934sx}.

In semi--elastic scattering, $p + p\rightarrow p+ X+ (\gamma \gamma)
\rightarrow p+ X + M$, the photon spectrum associated with the
interacting proton must be altered from the equivalent--photon
approximation for quarks to account for the proton structure.  To
accommodate the proton structure we use the modified
equivalent--photon approximation of \cite{Drees:1988pp}.

For elastic scattering, $p+ p \rightarrow p+ p+(\gamma\gamma)
\rightarrow p + p + M $,  both protons remain intact in the final
state.

\begin{figure}
\begin{center}
\epsfig{file=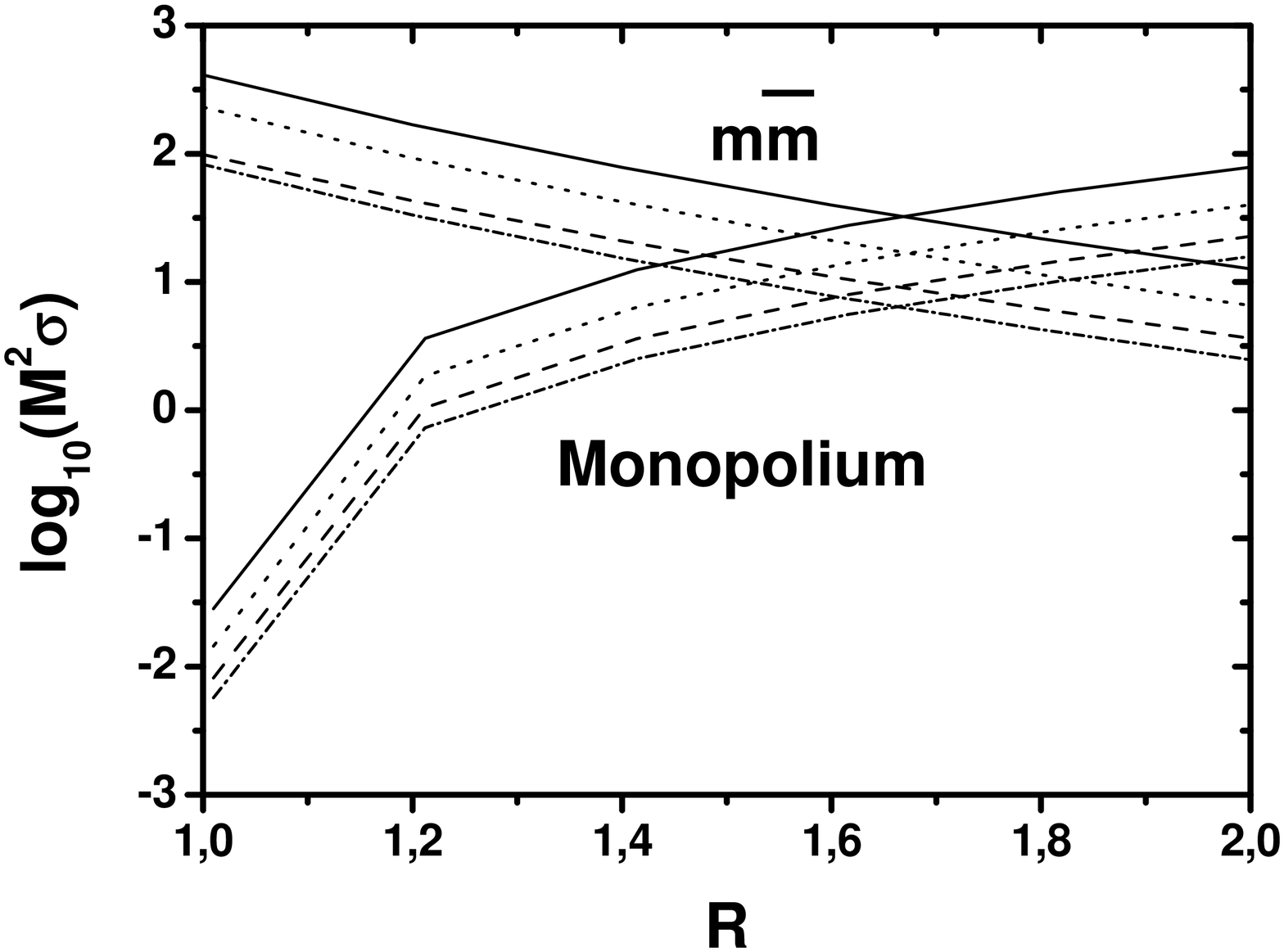,width=5.5cm,angle=270} \hspace{0.5cm}
\epsfig{file=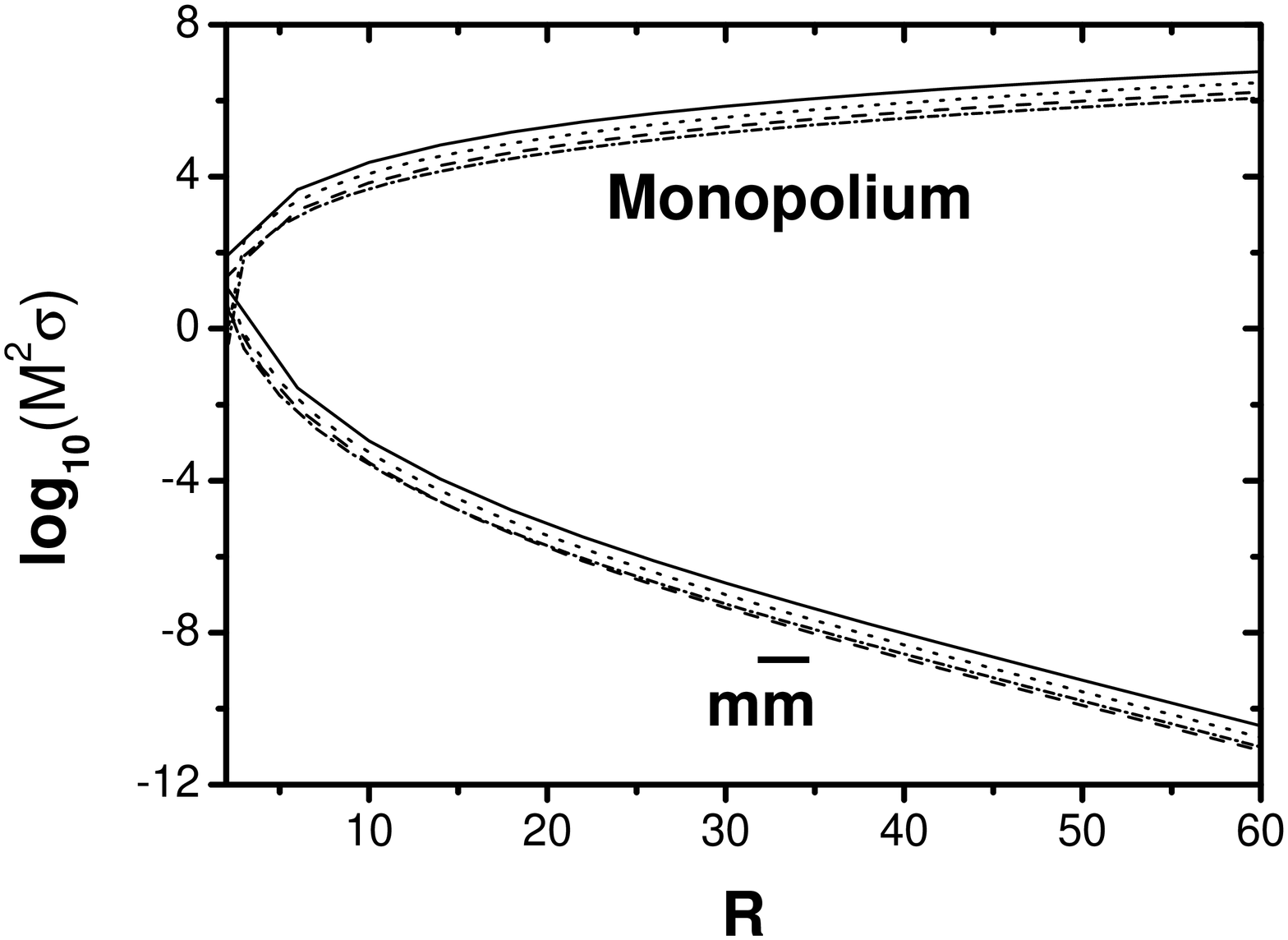,width=5.5cm,angle=270}
\end{center}
\caption{\small{The left figure shows  logarithmic plots of the $p
p$ production cross sections (in units of $M^2$) via photon fusion
for monopolium ($\Gamma_M =0.1 M$) and for monopole-antimonopole as
a function $R =2m/M$ for low binding. The figure on the right shows
logarithmic plots of the $p p$ production cross section (in units of
$M^2$) via photon fusion for monopolium ($\Gamma_M =0.1M$) and for
monopole-antimonopole as a function of $R$ for large binding. The
solid line represents the total cross section; the various
contributions are represented by: inelastic (dashed line);
semielastic (dotted line) and elastic (dot-dashed line). }}
\label{ppxsecR}\end{figure}

In Fig.\ref{ppxsecR} we show the comparison between
monopole-antimonopole and monopolium production cross sections using
$M$ as energy unit as a function of $R = 2m/M $, i.e., binding
energy $ |E_{binding}|/M = R - 1$. Two regimes are found:

\begin{itemize}
\item[ i)]the low binding regime $1 < R < 2$ where the $m\overline{m}$
production is dominant;

\item [ii)] the large binding regime $R > 2$ where monopolium production
is dominant and can be very large.

\end{itemize}

\begin{figure}
\begin{center}
\epsfig{file=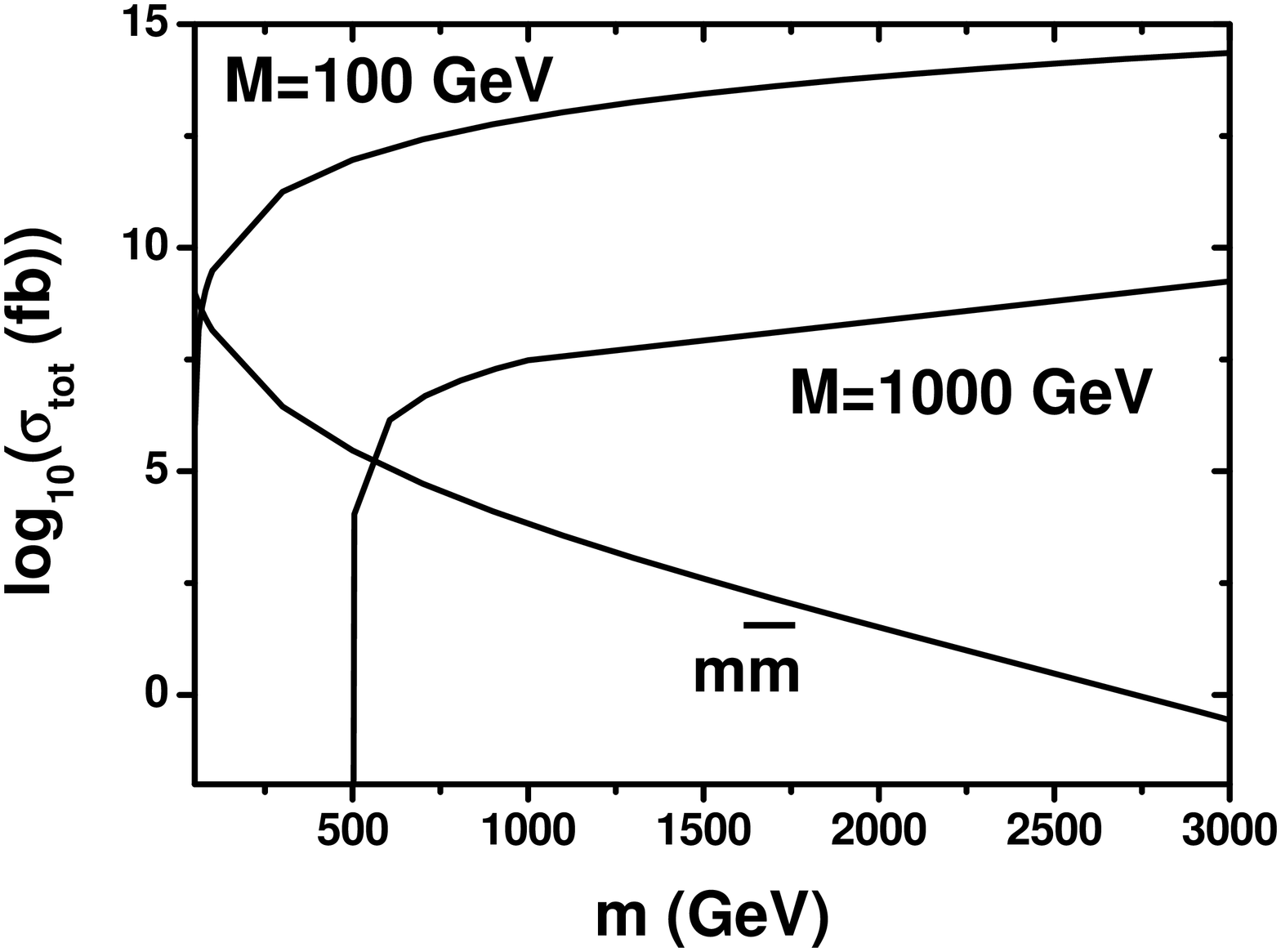,width=7.5cm,angle=270}
\end{center}
\caption{\small{We represent the logarithmic plot of the total $p p$
cross section from photon fusion in femtobarns as a function of
monopole mass, for two monopolium masses $M=100$ GeV and $M = 1000$
GeV and compare it with the corresponding monopole-antimonopole (m
\underline{m}) cross section.}} \label{ppxsecmass}\end{figure}

In Fig.\ref{ppxsecmass} we plot the logarithm of the total cross
section (in fb) for $pp$ production of monopolium as a function of
monopole mass, for two masses of monopolium (M=100 GeV and M= 1000
GeV), as well as the total cross section for monopole-antimonopole
production as a function of monopole mass. The two mass plot shows
the dependence with monopolium mass of the cross section. There is
always a threshold at $m = M/2$, i.e. for zero binding energy, where
the cross section vanishes, and then a rapid rise with monopole
mass. Thereafter the cross section grows in a softer manner,
although it should be realized that we deal with a log plot and
therefore the growth is not negligible. Finally, the smaller the
monopolium mass, the larger the cross section is.

The most spectacular signature of the curves is that for fixed
monopolium mass the cross section increases with monopole mass,
instead of decreasing, as happens in monopole-antimonopole
production, and the magnitude becomes enormous for very large
binding energy. Thus a strongly bound monopolium state would be an
ideal system to disentangle monopole dynamics.

\section{Conclusions}

We have carried out an investigation looking for hints of the so far
not seen monopoles. Our motivation has been, that of our previous
work \cite{Epele:2007ic}, namely that monopolium, if strongly bound,
is easier to produce than monopole-antimonopole pairs.

We have performed a calculation in which monopolium is produced via
the conventional monopole dynamics
\cite{Dougall:2007tt,Dougall:2007zz} used to study
monopole-antimonopole production by photon fusion at LHC energies.
Our calculation is parameterized in terms of three quantities: m,
the monopole mass, M the monopolium mass (or binding energy) and the
width $\Gamma_M$ which we simply take to be narrow compared with the
monopolium mass.

Our analysis distinguishes two clearly distinct regions associated
with the binding energy in monopolium. If the binding energy is
small compared to the monopole mass, the monopole-antimonopole
process will be dominant. On the contrary if it is comparable or
larger, the monopolium process is not only dominant but can be
extremely large.

The most favorable scenario, which agrees with that discussed in our
previous work, is a two energy scale scenario, whose

\begin{itemize}

\item[i)] low energy scale is governed by the monopolium mass,
$M$, reachable by LHC,

\item[ii)] and whose high energy scale is governed by the monopole
mass, $m$, which arises through the structure of monopolium, and
which could be larger than the energy reachable by LHC.
\end{itemize}

Under these circumstances the cross section as a function of the
monopole mass becomes sizeable.

An example, if monopole and monopolium have a mass of 1 TeV, at an
integrated luminosity of 100 $fb^{-1}$ at LHC, Dougall and Wick
predict about $10^6$ monopoles, while our calculation would produce
$10^8$ monopolia.

Since at present we cannot calculate the monopolium parameters, $M$
and $\Gamma_M$, the experimental endeavor is not easy. However, if
we extend the theory to incorporate the weak interaction
\cite{Ginzburg:1998vb,Ginzburg:1999ej} there are some features which
might simplify the task,

\begin{itemize}

\item[i)] the resonance peak of the monopolium can be found in three exit channels
2$\gamma$, $\gamma \, Z_0$ and $ 2\, Z_0$'s.

\item[ii)] monopolium can be produced in an excited state before
it annihilates, thus the annihilation process will be accompanied by
a Rydberg radiation spectrum;

\end{itemize}

The calculated values for the cross sections, corresponding to
reasonable monopolium mass scenarios, render our calculation sound
and this line of research worth pursuing.

\section*{Acknowledgement}
We thank the authors of JaxoDraw  for making drawing diagrams an
easy task \cite{Binosi:2003yf}. This work was done while one of us
(VV) was on a sabbatical from the University of Valencia at the
PH-TH at CERN, whose members he thanks for their hospitality. VV was
supported by MECyT-FPA2007 and by MEC-Movilidad PR2007-0048. LNE, HF
and CAGC were partially supported by CONICET and ANPCyT Argentina.

\end{document}